\documentclass[useAMS,usenatbib]{mn2e}
\pdfoutput=1
\usepackage{graphicx}
\usepackage{multirow}
\usepackage{multicol}
\usepackage{psfig}

\title[Morphological classification of galaxies]{Morphological classification of galaxies and its relation to physical properties}

\author[D. B. Wijesinghe, A. M. Hopkins, B. C. Kelly, N. Welikala and A. J. Connolly]
{D. B. Wijesinghe$^{1}$\thanks{E-mail:D.Wijesinghe@physics.usyd.edu.au},
A. M. Hopkins$^{2}$, B. C. Kelly$^{3,4}$, N. Welikala$^{5}$, A. J. Connolly$^{6}$\\
$^{1}$School of Physics, A28, University of Sydney, NSW, 2006, Australia\\
$^{2}$Anglo-Australian Observatory, P.O. Box 296, Epping, NSW, 1710, Australia\\
$^{3}$Harvard-Smithsonian Center for Astrophysics, 60 Garden St, Cambridge, MA 02138\\
$^{4}$Hubble Fellow\\
$^{5}$Laboratoire d'Astrophysique de Marseille, Technopôle de Marseille-Etoile, 38, rue Frédéric Joliot-Curie, 13388 Marseille Cedex 13, FRANCE\\
$^{6}$Department of Astronomy, University of Washington, Box 351580, Seattle, WA 98195-1580}
\begin{document}



\date{Accepted 2009 September 15. Received 2009 September 14; in original form 2009 20 August}

\pagerange{\pageref{firstpage}--\pageref{lastpage}} \pubyear{2009}

\maketitle

\label{firstpage}

\begin{abstract}
We extend a recently developed galaxy morphology classification method,
Quantitative Multiwavelength Morphology (QMM), to connect galaxy
morphologies to their underlying physical properties. The traditional
classification of galaxies approaches the problem separately
through either morphological classification or, in more recent times,
through analysis of physical properties. A combined approach has
significant potential in producing a consistent and accurate
classification scheme as well as shedding light on the origin
and evolution of galaxy morphology. Here we present an analysis
of a volume limited sample of 31703 galaxies from the fourth data
release of the Sloan Digital Sky Survey. We use an image analysis method called Pixel-z to extract the
underlying physical properties of the galaxies, which is then
quantified using the concentration, asymmetry and clumpiness (CAS) parameters.
The galaxies also have their multiwavelength
morphologies quantified using QMM, and these results are then related
to the distributed physical properties through a regression analysis.
We show that this method can be used to relate the spatial distribution
of physical properties with the morphological properties of galaxies.
\end{abstract}



\section{Introduction}
\label{int}
The Hubble tuning fork \citep{Hub:26} is one of the first and most widely used galaxy classification schemes. Even though it has provided many insights into the evolution of galaxies, their
morphologies and other physical properties, \citep{vdB:98} it remains subjective, requiring experts to manually classify galaxies,
and is not directly related to the physical properties of the galaxies. The use of one wavelength ($\approx$450nm) also restricts the Hubble scheme and studies such
as \citet{BW:91} and \citet{Jar:00} show clear examples of why a multiwavelength approach is vital for robust and thorough classification. Many galaxies have been discovered that simply 
do not fit into the Hubble scheme \citep{vdB:76,SB:79}. This is further exemplified at high redshift, where galaxies are at earlier stages of their evolution.
The inclination of galaxies also plays a vital part in the classification process, especially for those galaxies that fall into the spiral sequence.

Galaxy classifications need to be compact and physically motivated and the above limitations have motivated a multitude of classification schemes to achieve these goals while
addressing the various drawbacks of the Hubble tuning fork. De Vaucouleurs (1959) extended Hubble's scheme
by introducing more divisions within classes. For instance, spirals were divided into more refined classes based on the presence of bars and rings around the galaxy and the spiral arms were divided
into three new classes. The intrinsic basis of the Hubble system remains, though, limiting the utility of such schemes in progressing towards a connection between morphology
and the underlying physical processes. Simpler classification systems by \citet{M:58,M:59} using the central concentration of light and a system based solely on the stellar population of galaxies 
by \citet{MO:69} have been successful at providing high discrimination of galaxies between classes. \citet{KB:96} aimed to revise the
original classification of ellipticals by Hubble as this is mainly correlated to the inclination of galaxies rather than any intrinsic properties of the galaxies. These authors developed two main
classes for ellipticals and several branches within each type. As a subjective scheme however, the classes are not robust and many galaxies have properties that belong to multiple
classes or none at all. \citet{KB:96} attribute these inconsistencies to heterogeneous formation histories.

More recently automated classification schemes such as Artificial Neural Networks (ANNs) \citep{NRG:97a,Baz:00,Ode:02} have been used to accommodate the vast
quantities of galaxies that require classification and to a large extent ANNs eliminate any subjective bias in the classification process, as well as being very accurate \citep{Bal:04}.
ANNs cannot create a new classification system but rather
replicate visual classification and with much higher consistency. These systems require a ``test sample" previously classified by a human expert on which to base classifications 
which has the disadvantage of allowing the same human 
biases and flaws of the classification system to propagate. The use of Self Organising Maps by \citet{NRG:97b} however, eliminates the need for a test sample and any human 
influence in the classification process.

Photometric decomposition techniques which analyse the observed distributions of photometric intensity \citep{Sea:02,Pea:02} and Fourier analysis techniques that quantify luminosity distributions of
galaxies \citep{Ode:02,Trin:98} have had moderate success in differentiating galaxies into their respective classes.

Non-parametric approaches such as the CAS classification scheme have had success in objectively separating galaxies into Hubble's classes as well as being applicable at high redshifts (z=3)
\citep{Cons:03,Aea:96}. The scheme uses 3 properties, concentration, asymmetry and clumpiness, which quantify aspects of galaxy morphology. These quantities identify formation histories, merging 
activity
and areas of high star formation activity \citep{Cons:03}. This technique can be easily applied to the decomposition of the distribution of
other physical properties in galaxies. The inclusion of additional parameters such as a Gini Coefficient has been shown to produce more refined separations of galaxies but at the expense
of increasing the dimensionality of the classification scheme \citep{LPM:04,Lea:06}.

Classification of galaxies by physical properties alone has not been extensively carried out, although new techniques such as Pixel-z have emerged that enable the extraction of information about the
physical properties of galaxies by fitting spectral energy distribution (SED) templates derived from stellar population evolution \citep{BC:03,Con:03}.
Each pixel is fitted with templates, giving a localised analysis of the physical properties within the galaxy \citep{Wel:07,Wel:08,Wel:09}.  

Shapelet decomposition promises a new approach in the morphological
classification of galaxies. Shapelets are Gaussian-weighted Hermite polynomials 
\citep{Ref:03}.
They are also the eigenstates of the Quantum Harmonic Oscillator (QHO) Hamiltonian, and are thus well understood \citep{Ref:03}. 
They have been shown to be useful in image simulation \citep{Mea:04} and gravitational lensing measurements \citep{CR:02,RB:03}. Shapelets use all the information about the shape of a 
galaxy and
form a complete set thus making them an ideal candidate to be used in the morphological classification of galaxies \citep{KM:04}. Shapelets are a
central component in a new objectively developed and automated classification system known as the Quantitative Multiwavelength Morphology (QMM). QMM uses shapelets to decompose the galaxy images
and a Principal Component Analysis (PCA) to reduce the dimensionality of the data followed by a Mixture-of-Gaussian models to objectively identify particular morphological classes of galaxies
\citep{KM:04,KM:05}. Shapelets are not a compact form of classification and require a PCA to account for this. The technique uses galaxy
images in multiple filters, and currently images in the
Sloan Digital Sky Survey (SDSS) filter set (ugriz) has been used. \citet{KM:04} and \citet{KM:05}
show that this technique consistently reveals
previously established relationships such as that between Hubble type and colour, as well as broad connections between morphology and the physical properties of galaxies.

We aim to identify relationships between the physical properties, measured with Pixel-z and quantified using CAS, and the morphological properties quantified using QMM.


\section{Quantification of Physical Properties}
\label{data}
Our initial objective is to extract the physical properties from galaxy images and then quantify the spatial distributions of these physical properties.
The tools used for this process are Pixel-z \citep[see][]{Wel:07,Wel:08,Wel:09},
 for extracting the physical properties, and CAS \citep{Cons:03}, to quantify
their spatial distribution (the collective process will be known as
Pixel-z\,:\,CAS herein). 

These quantities are then compared with the results of the QMM analysis
of the same galaxy images, through a regression analysis, to analyse how
well QMM describes the spatial distribution of the physical properties in
galaxies. This indicates the extent to which we can use QMM to connect physical and morphological properties of galaxies, providing us with the possibility of 
developing a comprehensive galaxy classification scheme that incorporates both physical and morphological properties of galaxies. 

The data was obtained from the fourth data release (DR4) of the Sloan Digital Sky Survey (SDSS). The SDSS used a dedicated 2.5m 
telescope located at Apache Point Observatory in New Mexico, USA 
together with a 142 megapixel camera in drift-scan mode to obtain images and spectroscopy over about a quarter of the sky \citep[for more details, see][]{Yokea:00,Hogea:01,Smtea:02,
Subea:02,Pirea:03,Izcea:04,Tkrea:06}

\subsection{Parameterization of Pixel-z Output}
Using the Pixel-z output from \citet{Wel:08} we quantified the distribution of the physical properties in galaxies through the CAS technique developed by \citet{Cons:03}.
The CAS analysis uses the parameters concentration, asymmetry
and clumpiness \citep[C, A and S,][]{Cons:03}. We calculated these three parameters for four physical 
properties, age, star formation rate, colour excess and metallicity, each of
which were obtained from the Pixel-z analysis of \citet{Wel:08}. The same process was carried out for the
$r$-band (616.5nm) image of the galaxy.

From the $r$-band CAS parameters we can contrast the relationships between QMM and physical properties 
with that between QMM and the photometric distribution for the galaxies. The $r$-band is convenient for this process as it lies in the middle of the SDSS filters and typically shows a 
high signal to noise ratio. 

The concentration parameter is calculated from the ratio of the radii containing 20\% and 80\% of the total flux. Asymmetry is derived from subtracting the image of a galaxy 
rotated by 180$\,^\circ$ from its original image. The clumpiness is quantified through the ratio of the flux in high spatial frequency structures within a galaxy to its total flux.
These CAS parameters provide a morphological description of the physical properties of galaxies by quantifying their distributions derived from Pixel-z.

\section{Quantitative Multiwavelength Morphology}
QMM is a morphological classification method developed by \citet{KM:04, KM:05}, using galaxy images observed through
the five SDSS filters ($u$, $g$, $r$, $i$, $z$). Our implementation of QMM involves two steps, shapelet decomposition and a Principal Components Analysis (PCA). 
The shapelet decomposition breaks down the image of the 
galaxy represented in all 5 filters into a set of mathematical functions which are quantified by a set of coefficients. To adequately reconstruct galaxy images the number of coefficients
can be in the hundreds, so PCA is used to reduce the number of coefficients to a more manageable size. The outputs of the QMM analysis are the final PCA components,
which encompass the majority of the variance in the data.

For the shapelet decomposition 37876 galaxy images from the Pixel-z:CAS analysis were available, comprised of the volume-limited sample analyzed by \citet{Wel:08, Wel:09}.
The resulting data set excluded 432 galaxies, the images for which
contained numerous zero flux values that could not be processed in the PCA analysis.
The sample was further restricted with an 11$\sigma$ cut on the variances in each dimension in order to ensure that the PCA was not compromised by outlying data points.
The final sample for the QMM analysis consisted of 31703 galaxies with images from all five filters and common to both the Pixel-z\,:\,CAS and QMM analysis.    

The galaxies in the sample span a redshift range of $0.00278 < z < 0.231$. 
The analysis was first carried out over the entire redshift range.The analysis was then independently applied to four redshift
bin sub-samples containing approximately equal numbers of galaxies.
This approach allows us to identify possible biases due to the coarser
spatial sampling of galaxies at progressively higher redshift, imposed
by a fixed observational pixel scale, and identifying whether
it is necessary to artificially redshift the galaxy images to mimic
a common spatial sampling.

By comparing the results from QMM with those of Pixel-z\,:\,CAS we can explore how QMM enables us
to connect galaxy morphology with the underlying physical properties. The colour (or flux ratio) information
encoded in the multiwavelength images used for this technique provides the key, as galaxy colours are a consequence of the combination of stellar evolutionary processes and 
multiple stellar populations. This suggests that there
should be a direct connection between a quantitative morphology derived from multiwavelength images and the underlying properties of the stellar population within the galaxies. Therefore, we aim to 
determine the extent to which QMM is a reliable descriptor of the underlying physical properties of galaxies by comparing it with the results of the Pixel-z\,:\,CAS analysis.




\section{Results}

\subsection{Pixel-z\,:\,CAS}
The CAS analysis provides morphology descriptions for
the four sets of physical parameters from Pixel-z: age,
star formation rate, colour excess and metallicity. For each of these the 
concentration, asymmetry and clumpiness parameters were measured.

\begin{figure*}
\centerline{\includegraphics[width=80mm, height=70mm]{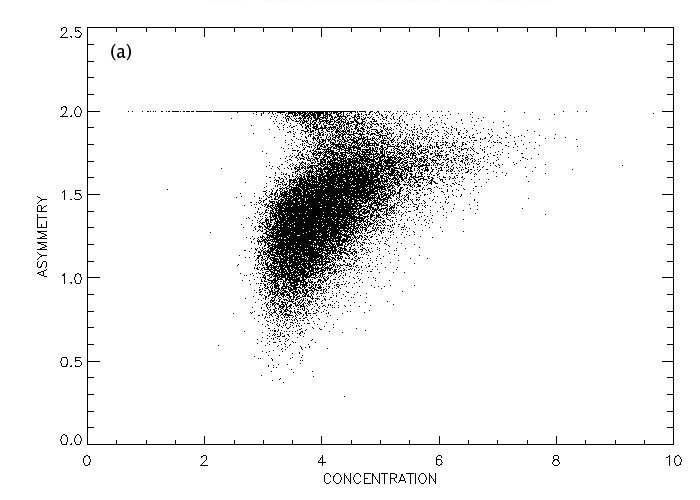}}
\centerline{\includegraphics[width=80mm, height=70mm]{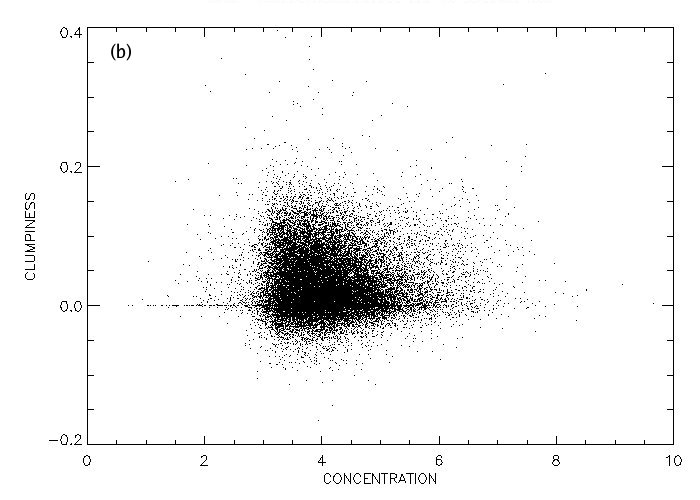}\includegraphics[width=80mm, height=70mm]{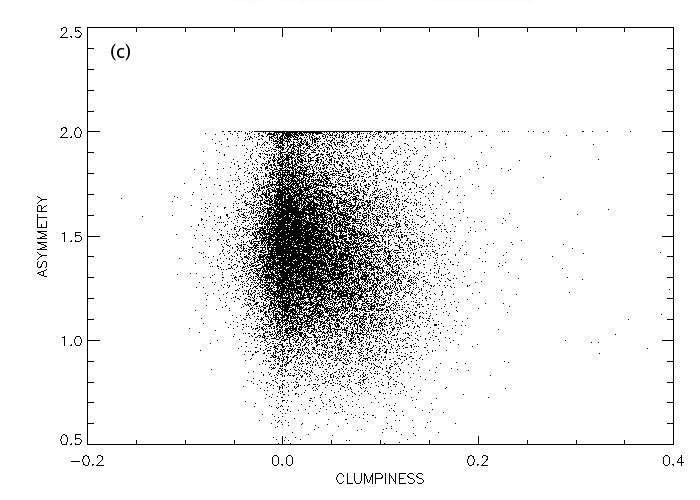}}
\caption{Relationships between the three CAS parameters for the spatial distribution of star formation rate as inferred
by Pixel-z, with (a) concentration against asymmetry, (b) concentration against clumpiness
and (c) clumpiness against asymmetry.}
\label{fig:sfr}
\end{figure*}



Our results for the distribution of the star formation rate of galaxies in CAS space (Figure~\ref{fig:sfr}) are not consistent with those found by
Conselice (2003) for the photometric images. This is not unexpected, as this is the first time that CAS has been applied to a distribution
of physical rather than photometric properties. The ranges spanned by our CAS values are different, extending
over a much larger range than those of \citet{Cons:03}. The clumpiness parameter in particular, (Figures 1b, 1c), shows
a more restricted distribution than in \citet{Cons:03}, with a number of systems having clumpiness very close to zero. The asymmetry
parameter shows a maximum value of 2, a consequence of the mathematical definition of asymmetry.

Figure~\ref{fig:sfr} (a) shows that at lower concentrations there is no relationship between concentration and asymmetry. As we move to higher concentrations the 
asymmetries fall between 1.5 and the highest possible value of 2. At high asymmetries the star formation activity is occurring only in highly localised regions of the galaxy. 
The proportion of light emanating from these regions within the galaxy compared to the central bulge is smaller than in a galaxy with a low asymmetry. This leads to a high
concentration index.

Clumpiness is not correlated with concentration for SFR at any value.
Figure~\ref{fig:sfr} (b) shows uniform distribution above Concentration\,=\,3 and Clumpiness\,=\,0.
A possible explanation for this could be that the clumpiness parameter takes into account the galaxy centers as well as
the regions outside the center. The clumpiness of the galactic center would be expected to correlate positively with the concentration. The clumpiness outside the
central regions would be expected to correlate negatively with the concentration, as the more clumpy the outer regions, the more localised the SFR activity. This leads to a 
higher concentration index as in the case of Figure~\ref{fig:sfr} (a). When the clumpiness of both these regions are taken into consideration, we are more likely 
to see a large scatter. The lack of correlation between asymmetry and clumpiness seen in Figure~\ref{fig:sfr} (c), has a similar 
explanation to that of Figure~\ref{fig:sfr} (b).

The CAS distributions for other Pixel-z physical properties also differ from those for the r-band light distribution of galaxies from \citet{Cons:03}.
Simply by analysing these relationships in CAS space it is apparent that the distribution of physical parameters in galaxies does not mimic its photometric morphology in a trivial fashion.
This highlights the fact that a simplistic approach to relating the
single-filter photometric morphology (parameterised by CAS) with the
distribution of physical properties within galaxies does not show
any obvious connection. This in turn supports our motivation for
using multiwavelength photometric information, through QMM,
to relate galaxy morphology to the underlying physical properties.




The use of CAS in this investigation is not as a morphological
indicator directly, but instead as a convenient technique for
parameterising the distributions of Pixel-z physical parameters,
to allow a comparison with the QMM results from the shapelet decomposition.


It is possible to use the differences in the physical parameters to refine the boundaries for the different galaxy classes
set by \citet{Cons:03}. For instance, \citet{Cons:03} set the CAS ranges for ellipticals to be, concentration in the range 4.4$\pm$0.3, 
asymmetry in the range 0.02$\pm$0.02 and clumpiness in the range 0.00$\pm$0.04. As we use larger numbers of galaxies and at higher redshifts, these values will undoubtedly 
vary. If we use the four physical properties and their CAS boundaries for different galaxy types, it may be possible to set a more consistent limit for
each class of galaxies. This extension of CAS to morphology classification based on physical properties could be a productive area for further investigation, but is beyond
the scope of this paper.

\begin{figure*}
\centerline{\includegraphics[width=80mm, height=70mm]{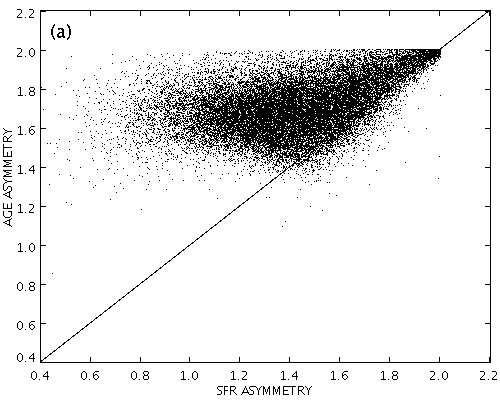}
\includegraphics[width=80mm, height=70mm]{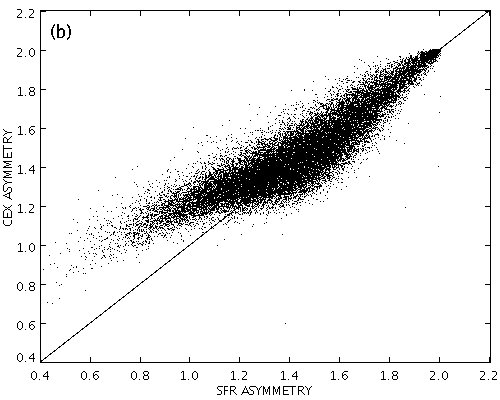}}
\centerline{\includegraphics[width=80mm, height=70mm]{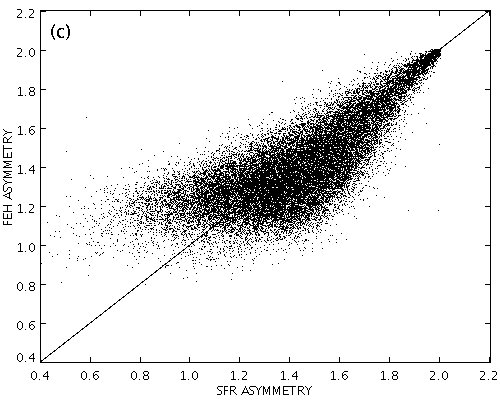}
\includegraphics[width=80mm, height=70mm]{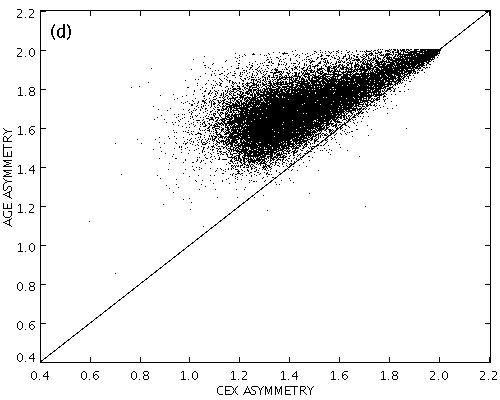}}
\centerline{\includegraphics[width=80mm, height=70mm]{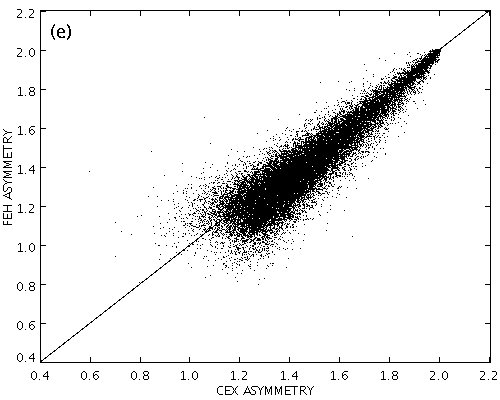}
\includegraphics[width=80mm, height=70mm]{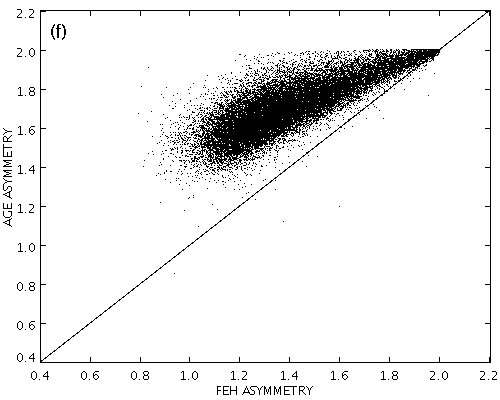}}
\caption{Comparison of all physical properties against each other in the asymmetry parameter. Distributions are in the same ranges for ease of comparison and the 
line crossing each plot is the one to one relation. Figures (a), (b) and (c) compare the asymmetry of the SFR against the asymmetry of age, colour excess and metallicity 
respectively. Figures (d) and (e) plot the asymmetry of colour excess with the asymmetry of age and metallicity while figure (f) compares the asymmetry of metallicity with age. 
The asymmetry of SFR, colour excess and metallicity follow each other to a large extent. \label{fig:a_vs_a}}
\label{fig:asym}
\end{figure*}

Figure~\ref{fig:asym} shows the relationship between the physical properties with regards to the asymmetry parameter. They allow for an understanding of how these
properties affect each other and how they can be used to define classification criteria. Figure~\ref{fig:asym} (e) shows the best agreement out of all the tested 
parameters as well as the strongest correlation. Both colour excess and metallicity agree with the SFR, but the correlation is weaker than that between colour excess and 
metallicity. The asymmetry in age does not match with the asymmetry in any of the other parameters but the tightness in the relationship between age and metallicity 
(Figure~\ref{fig:asym} (f)) has potential in being exploited to make finer cuts in the morphological classification. Strong relationships between these properties
allow for establishing criteria for better galaxy classification than if used on their own.

The spread of galaxies at high asymmetries in Figure~\ref{fig:asym} is highly constrained. High asymmetries indicate that the distribution of the
various physical properties will be very ``patchy''. As a result, these properties will be found in highly localised regions of each galaxy. As these properties 
have some relation to one another, it is likely that they will all be found in the same localised regions, resulting in highly asymmetric galaxies having high asymmetries for the measured attributes.
This also explains the large distributions of galaxies at low asymmetries as the distribution of physical properties within the galaxy will not be ``patchy'' allowing for the
bigger spread in the physical properties across the galaxy. This spread will be largely independent of the  other physical properties leading to lower correlations at low
asymmetries.

There is a limit of 2 for all the asymmetry parameters in Figure~\ref{fig:asym}. This is an artifact of the definition of
asymmetry. Asymmetry is defined as the absolute value of the difference in flux between the original image and the same image rotated by $180\,^{\circ}$,
divided by the flux of the original image. The maximum possible value for asymmetry that can be attained with this definition of asymmetry is two.

\subsection{QMM - Principal Components Analysis}

Performing the shapelet decomposition on the galaxy images using
a shapelet order of 12 resulted in 91 coefficients per filter, making 455 coefficients per galaxy. 
A lower number of coefficients is insufficient to reproduce galaxy images reliably and even though a higher number of coefficients would have given a more accurate 
representation of the galaxy images it would have been too large to have been effectively reduced by the PCA. 

A PCA was used to reduce the dimensionality of the results
of the shapelet decomposition \citep{Kar:47,Lov:78}. 
To carry out the PCA we first used the same procedure described in \citet{KM:05} to calculate the shapelet coefficients, with two notable differences. As mentioned above, we do not artificially
re-sample the images to
the same redshift. Also, we use a Singular Value Decomposition (SVD) technique similar to the one described in \citet{BHW:04} to decompose the shapelet coefficients. The SVD accounts
for the loss of orthogonality in the shapelets that result from the pixelization of the images. However, unlike \citet{BHW:04}, the number of non-zero singular values is chosen in an objective
manner. In our work, we chose the number of non-zero singular values to minimize the estimated squared error between the true galaxy image and the galaxy image reconstructed from the shapelet
coefficients. We use Stein's Unbiased Risk Estimation to estimate this error \citep{Stn:81}, which is easily calculated as a function of the number of non-zero singular values.

The first component contains the majority of the variance ($\approx$75\%) in the data set. Even though the subsequent components contain a much 
lower fraction of the variance they are nevertheless important. The number of components was limited to 8, which contains 92.3\%  of the total variance and is 
sufficient for our purposes.


\begin{figure*}
\centerline{\includegraphics[width=80mm, height=70mm]{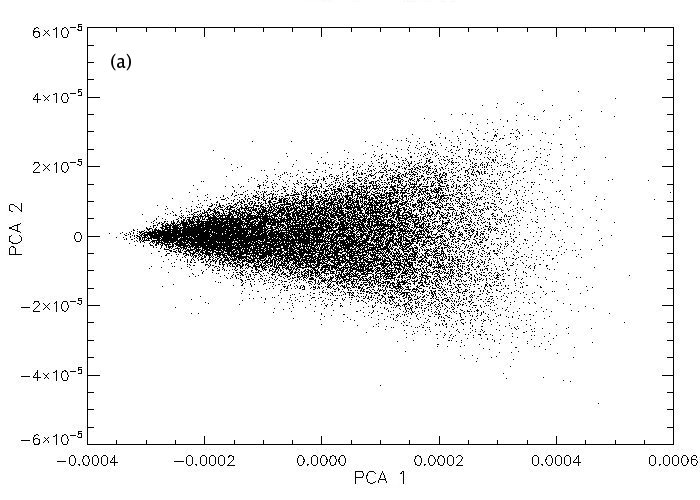}}
\centerline{\includegraphics[width=80mm, height=70mm]{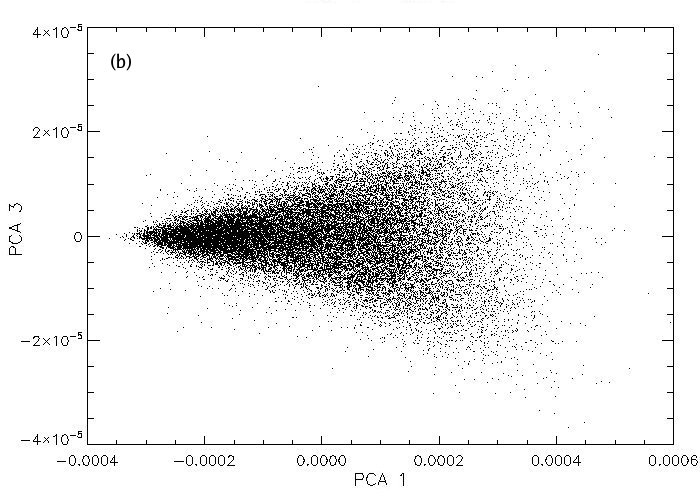}
\includegraphics[width=80mm, height=70mm]{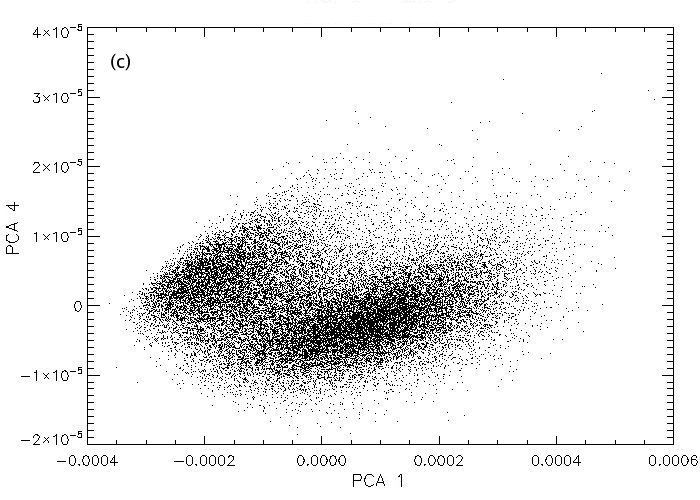}}
\caption{Graphs for the first principal component against the next three. Note the different divisions shown. Components 2 and 3 have a diverging pattern while the 4th component has 
two clearly distinct groups. These results reproduce well those of Kelly \& McKay (2005), but the separation for PCA 4 shown above was found for PCA 2 by \citet{KM:05}.
\label{fig:pca_vs_pca}}
\end{figure*}


Discrimination between galaxy populations is not apparent in the first two panels of Figure 3, but becomes clear in the third panel, plotting PCA 1 against PCA 4.
This bimodality in galaxy morphology has long been known to exist and it is related to the two broad morphological
classes of galaxies, early and late types.
\citet{KM:05} detected this same bimodality, between their PCA\,1
against PCA\, 2. The fact that this bimodality is seen here in a higher order PCA component could be a result of many factors. \citet{KM:05}
carried out the principal component analysis using a sum-of-squares matrix while we used a covariance matrix. Also, even though we used a sigma cut to limit any extreme
outliers, there may still be fewer distant outliers that can significantly affect the results of PCA. Thus the location of the bimodality in PCA space does not reveal any
fundamental property of galaxies as PCA is simply another way of representing the distribution of galaxies. The other separations in PCA space are
very similar to the separations obtained by \citet{KM:05}.

Similarly, the galaxy types (or combination of them) that each PCA component represents will also depend on the above factors. To objectively identify particular classes contained within the PCAs,
for a particular sample, one must apply a Mixture-of-Gaussian methods \citep{KM:04,KM:05} but this falls outside the aims of this experiment.

\begin{figure*}
\centerline{\includegraphics[width=80mm, height=70mm]{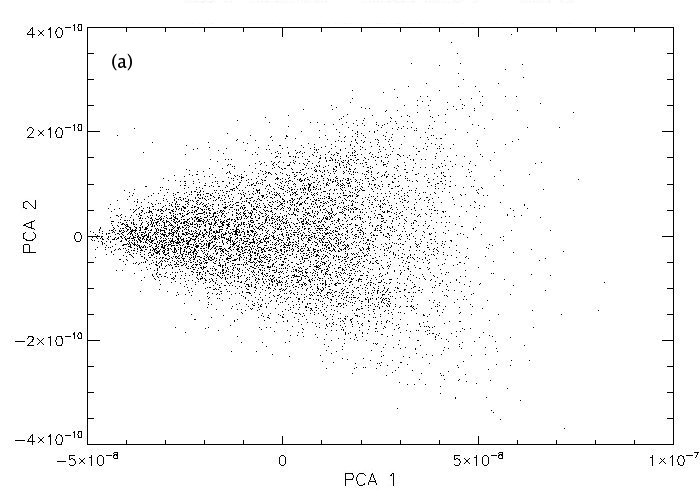}
\includegraphics[width=80mm, height=70mm]{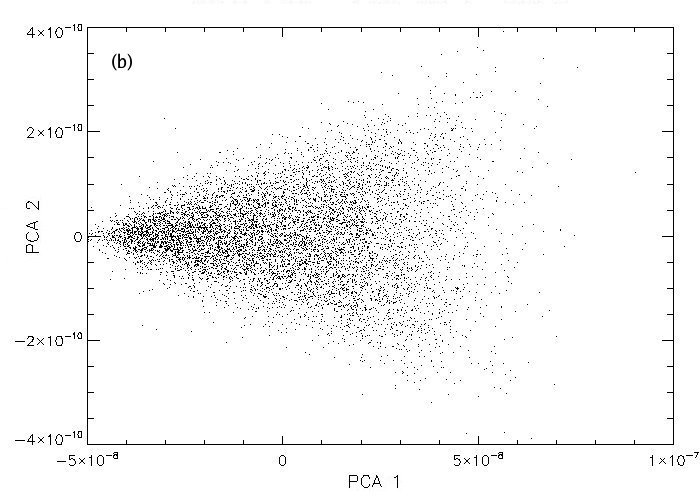}}
\centerline{\includegraphics[width=80mm, height=70mm]{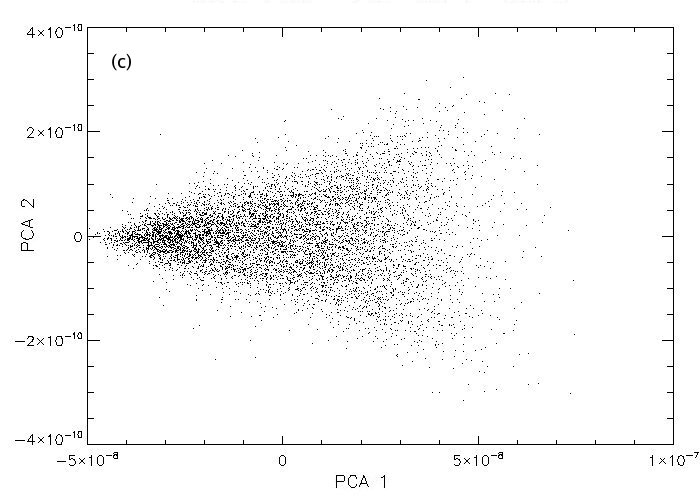}
\includegraphics[width=80mm, height=70mm]{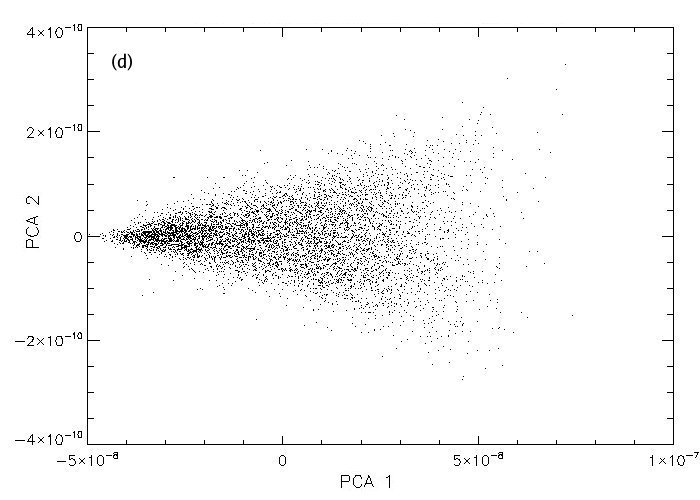}}
\caption{Principal component analysis for galaxies in the redshift bins. The distributions are very similar and a statistical analysis of these similarities is shown
in Table 2. \label{fig:pca_bins}}
\end{figure*}

This analysis was reproduced for the sub-samples
split into redshift bins. The distributions of galaxies in PCA space in each of the redshift bins are similar to the analysis with the full sample.
Furthermore, these distributions are consistent with each other (Figure~\ref{fig:pca_bins}), as well as those of \citet{KM:05}. 
These distributions of galaxies in PCA space in each of the redshift bins fall within
very similar ranges. These similarities among the different redshift bins allow us to conclude that our results are not strongly dependent on the treatment of the sample as a 
whole despite the range of redshift. Thus there is no need to simulate a common redshift, through artificially redshifting the galaxy images by reducing their resolution, 
as implemented by \citet{KM:05}. The principal components for the redshift bin subsamples are also correlated against the Pixel-z:CAS measurements below, to identify
any potential statistical differences.

As with the Pixel-z\,:\,CAS analysis, our objective is not to define a classification system through the distribution of the QMM parameters of
galaxies. Rather we aim to identify relationships between the two methods to identify the connection between the multiwavelength morphology and
the spatial distribution of the underlying physical processes. The next step in our analysis is to quantify the relationships between the
two methods.

\section{Correlation Analysis}
Morphological classification alone provides a limited approach for understanding the properties and evolution of galaxies. Using
just the physical properties may be more straightforward but these schemes do not address the origin or evolution of galaxy morphology. However, the multiwavelength nature of 
the recently developed QMM classification method provides an opportunity to connect morphological classification with underlying physical parameters. The colour information
encoded in the multiwavelength images provides the key, as galaxy colours are a consequence of the combination of stellar evolutionary processes and multiple stellar 
populations. This suggests there should be a direct connection between a quantitative morphology derived from multiwavelength images and the underlying properties of the 
stellar population within the galaxies. 

We have used Pixel-z and CAS as a way of quantifying several physical properties of galaxies so that they can be correlated with the QMM method, in order to identify how well 
QMM can represent the distribution of physical properties in galaxies. In order to connect the two methods, we carry out a regression analysis 
by performing  a multiple linear regression fit
to identify the extent to which QMM correlates with spatial distributions of physical properties in galaxies represented through the Pixel-z\,:\,CAS method. 
In an ideal case we would find strong correlations between the tested parameters, but due to the large scatter seen in the above figures this is unlikely. 
For this reason we focus on the relative correlations between the CAS parameters measured for the physical properties and the $r$-band to observe whether
any physical properties consistently show correlation coefficients higher than the $r$-band CAS parameters. 

To do this we carry out a regression analysis of 
the PCA components for each physical property against the CAS analysis of the $r$-band and $u$-band light distribution of the galaxy. The results are shown in 
Table~\ref{table:corr} for the full sample and Table~\ref{table:bins} for each of the smaller redshift range subsamples. The values in Table~\ref{table:corr} and 
Table~\ref{table:bins} show the correlation coefficients between the distribution of each physical property in a galaxy (described by 8 principal components) and the three 
CAS parameters. They also show the correlation coefficients between the $r$-band and $u$-band light distribution (also described by 8 principal components) and the
CAS parameters.

\begin{table*}
\caption{Correlation coefficients between the CAS parameters and the distribution of physical properties in the galaxy (including the $r$-band and $u$-band light
distributions.\label{table:corr}}
\centering
\begin{tabular}{ccccccccc}
\hline\hline \\
\multirow{2}{*}{CAS Parameters} & \multicolumn{4}{c}{Physical Properties} & \multicolumn{1}{c}{ } & \multicolumn{2}{c}{Light Profiles} \\ 
\cline{2-5} \cline{7-8} \\
 & Age & SFR & Colour Excess & Metallicity & & $r$-band & $u$-band \\ [0.5ex]
\hline \\
Concentration & 0.04 & 0.01 & 0.02 & 0.03 & & 0.20 & 0.04\\
Asymmetry & 0.18 & 0.24 & 0.21 & 0.20 & & 0.27 & 0.21\\
Clumpiness & 0.30 & 0.46 & 0.43 & 0.46 & & 0.28 & 0.40\\ [1ex]
\hline
\end{tabular}
\end{table*}



For this analysis, the CAS parameterization of the $r$-band images provides a convenient proxy for a single-filter derived morphology. Comparing the QMM results to the $r$-band
CAS parameters, as well as to the Pixel-z\,:\,CAS output (physical properties), allows us to establish whether the QMM measurements contain more information about
the distribution of physical properties within a galaxy than about the simple distribution of light.

If the correlation of the PCA components is higher for the CAS analysis of the physical properties compared to the $r$-band light 
profile, QMM would seem to be better at representing the spatial distribution of physical properties in galaxies than just their simple $r$-band light distribution.
The regression analysis was carried out on the entire galaxy sample as well as independently for the redshift subsamples, in order to identify any systematic effect due to varying redshifts.
The tables report the correlation coefficients between the PCA output and each of the CAS parameters for each property, and for the $r$-band light distribution.

There is essentially no correlation between the concentration of the physical properties and the principal components, particularly when these values are compared with the 
correlation for the $r$-band light distribution. The asymmetry of physical parameters has a slightly higher correlation with the output of QMM but even this is weaker than the 
correlation between QMM and the asymmetry of the $r$-band light distribution. Thus, the QMM morphology cannot describe the concentration and asymmetry of physical parameters better 
than it can describe the $r$-band light distribution. 

The correlations between the physical properties in the clumpiness parameter
and the QMM results are much higher than the correlation between QMM and the $r$-band light distribution. This indicates that the results of QMM 
provide a more accurate representation of the underlying physical parameter distribution, for the clumpiness parameter, than of the $r$-band light distribution of the galaxy. Thus QMM, 
a morphology indicator, can represent the clumpiness measurement of physical properties of galaxies better than the single-filter photometric light distribution.

However, there is a possibility that the higher correlation we find between the clumpiness parameter and QMM output could be the result of noise in the $u$-band data from the SDSS images which was
also processed by the QMM technique. This noise would make the image appear clumpy hence providing a better correlation with the clumpiness parameter. On carrying out the correlation analysis with the
QMM output and the $u$-band light distribution it is apparent that clumpiness of the star formation rate, colour excess and metallicity still have a higher correlation with the QMM output than the
$u$-band output. Thus we can conclude that the correlation between the QMM output and the clumpiness of physical properties is not due to any noise in the $u$-band images and that QMM actually does
represent the clumpiness of physical properties.

Following the analysis for the complete sample, the subsamples in redshift bins were analyzed independently, with similar results,
confirming that there is little or no systematic effect based on redshift (see Table~\ref{table:bins}).

   \begin{table*}
   \caption{CAS parameter values for the physical properties across each redshift bin as well as the $r$-band light distribution.\label{table:bins}}
   \begin{center}
   \begin{tabular}{ c c  c c c c  c c c }
   \hline \\
   \multirow{2}{*}{Redshift Range} & \multirow{2}{*}{CAS Parameters} & \multicolumn{4}{ c }{Physical Properties} & \multicolumn{1}{c}{ } & \multicolumn{2}{ c }{Light Profiles} \\ [0.5ex]
   \cline{3-6} \cline{8-9} \\ [-0.5ex]
   &  & Age & SFR & Colour Excess & Metallicity  & &  $r$-band & $u$-band \\
   \hline \\ [-0.75ex]
   \multirow{3}{*}{0.00278$\le{z}<$0.066}
   & Concentration & 0.05 & 0.16 & 0.04 & 0.04 & & 0.23 & 0.05\\
   & Asymmetry & 0.20 & 0.27 & 0.23 & 0.23 & & 0.30 & 0.21\\
   & Clumpiness & 0.26 & 0.46 & 0.42 & 0.44 & & 0.24 & 0.37\\ [0.5ex]
   \hline
   \hline \\ [-0.75ex]
   \multirow{3}{*}{0.067$\le{z}<$0.075}
   & Concentration & 0.04 & 0.02 & 0.02 & 0.03 & & 0.20 & 0.05\\
   & Asymmetry & 0.20 & 0.27 & 0.21 & 0.20 & & 0.28 & 0.21\\
   & Clumpiness & 0.31 & 0.46 & 0.42 & 0.46 & & 0.28 & 0.40\\ [0.5ex]
   \hline
   \hline \\ [-0.75ex]
   \multirow{3}{*}{0.076$\le{z}<$0.083}
   & Concentration & 0.22 & 0.02 & 0.20 & 0.03 & & 0.19 & 0.06\\
   & Asymmetry & 0.18 & 0.24 & 0.20 & 0.18 & & 0.26 & 0.21\\
   & Clumpiness & 0.32 & 0.47 & 0.43 & 0.47 & & 0.36 & 0.39\\ [0.5ex]
   \hline
   \hline \\ [-0.75ex]
   \multirow{3}{*}{0.084$\le{z}<$0.231}
   & Concentration & 0.19 & 0.11 & 0.02 & 0.26 & & 0.19 & 0.06\\
   & Asymmetry & 0.17 & 0.24 & 0.20 & 0.19 & & 0.26 & 0.28\\
   & Clumpiness & 0.32 & 0.45 & 0.42 & 0.46 & & 0.34 & 0.49\\ [0.5ex]
   \hline
   \end{tabular}
   \end{center}
   \end{table*}

For the separate redshift bins, all show similar patterns, and all are in agreement with the results for the complete sample.
These results confirm that QMM is a better descriptor of the clumpiness of the physical properties compared to the $r$-band 
light distribution of the galaxies. For the concentration and asymmetry parameters
QMM cannot describe the distribution of the physical properties any better than it describes the $r$-band light distribution of the galaxies.

An important finding here is that the relative correlations between physical properties and the $r$-band light distributions are consistent across all redshift ranges.
The actual values between different redshifts vary, but they are still consistent with the findings from the complete sample. These fluctuations can be 
attributed to the fact that the redshift ranges themselves are not even. The two middle bins are particularly small compared to the two redshift ranges at the edges of the 
sample. The ranges were chosen such that the number of galaxies in each is equal, so as to ensure that the principal component analysis is not biased by different sample sizes.
Furthermore, the fluctuations do not display a trend with higher redshifts, indicating that there is no real systematic effect due to redshift. 
Thus we conclude that QMM seems to provide a useful approach toward connecting galaxy morphologies to the morphologies of the physical parameters, 
and is independent of systematic effects due to redshift, over the small range considered here.

We stress that the correlation coefficients should not be taken as indicating real correlations. The correlation coefficients in all cases are 
too small to indicate any strong correlations between the tested parameters. Our goal is to compare the strength of the coefficients in relation to the $r$ and $u$-bands 
and identify those CAS parameters for the physical properties that have consistently higher correlations, if any, than in the $r$ and $u$ bands.

\section{Discussion}
We have two main findings. The first is that the correlation between the results of QMM and the clumpiness of physical properties of Pixel-z\,:\,CAS is 
significantly higher than the correlation with the clumpiness of $r$-band or $u$-band light distributions.
For the concentration and asymmetry parameters, the correlation between QMM and the physical properties
is much lower than the correlation with the $r$-band light distribution.

It is expected that the QMM results will be better correlated with the $r$-band light distribution of the galaxy as compared to the derived 
physical quantities as the QMM results are calculated from the light distribution of the galaxy. Thus it is indeed significant that the QMM results are better 
correlated with the clumpiness of the derived physical properties than with the clumpiness of the $r$-band light distribution. 

The lack of correlation in the concentration and asymmetry parameters is perhaps not completely unexpected. Neither of these parameters are particularly effective 
at describing the ``patchy'' nature of the distribution of physical
properties in galaxies. The Pixel-z decomposition of the spatial distribution of various physical properties in galaxies \citep[as seen in][]{Wel:08,Wel:09},
is an excellent example of this type of distribution. 
Unlike the light profile of galaxies, physical properties are not necessarily centrally concentrated. As a consequence, it seems
the concentration parameter is unlikely to be a useful descriptor for 
characterising the distribution of physical properties of galaxies, particularly compared to the optical light distribution.

The asymmetry parameter may also not be effective at describing the ``patchy'' distribution of physical properties, and it is further limited given that a center of rotation needs to be identified.
The brightest pixel was used as the center of rotation, which is a reasonable approximation for the optical light distribution, but the light and physical properties do not share
well correlated distributions and the high spatial frequency nature of the physical properties means that asymmetry is probably not the best approach for quantifying these.
Moreover, such ``patchy'' distributions are inherently asymmetric, and many systems would show an asymmetry close to the maximum possible asymmetry value (Asymmetry=2).
For these reasons asymmetry, too, seems not to be a
particularly useful parameter in describing the distribution of physical properties in galaxies.

Finally, there is the clumpiness parameter, which would clearly seem to be a more useful statistic in quantifying these ``patchy'' distributions of physical 
properties of galaxies than the previous two, as it was developed to measure this type of distribution. 
The correlation coefficients measured do not indicate a real
correlation, but we serve to distinguish the relative correlation strength of QMM with the pixel-z\,:\,CAS measurements compared to $r$ and $u$-bands. 
The range of values of clumpiness is more likely to be encoding much more physical information about galaxies, rather than the less useful and potentially biased estimates for
concentration and asymmetry. The QMM morphology should also be sensitive to the ``patchy'' distributions of pixel colours that underlies the clumpy nature of the physical properties.
This could well be the reason for the higher correlation values with QMM seen for clumpiness compared to the concentration and asymmetry as well as the clumpiness of the
$r$-band light distribution. We also confirmed that this clumpiness is not due to any noise in the $u$-band images by showing that the correlation
between the clumpiness parameters and QMM output is higher than that between the $u$-band and the QMM output. This was true for all redshift bins as well. 
Using other approaches towards quantifying clumpy structure may be helpful in confirming this.

\section{Summary}
We have compared two separate techniques for analysing galaxies, one using a purely physical approach (Pixel-z\,:\,CAS)
and the other using a purely morphological approach (QMM). The purely physical approach uses the Pixel-z method to infer
the distribution of four physical properties (age, star formation rate, colour excess and metallicity), within galaxies.
These distributions were then quantified using the CAS method where each physical property was assigned a value for its concentration, asymmetry and clumpiness. 
This is the first application of CAS to distributed physical properties within galaxies.
The Pixel-z\,:\,CAS procedure was also applied to the $r$-band and $u$-band light distribution. We did not discover a trivial relationship between the 
optical light distribution of galaxies and the distribution of the 
physical properties in CAS space. This places further significance on our objective of using QMM to connect the multiwavelength photometric morphology with the underlying physical properties. 
This is not unexpected as CAS is able to be applied only to single-filter imaging, rather than a full multiwavelength morphology.

We also analyzed the morphology of the galaxies using QMM where the images of the galaxies in five filters were decomposed using shapelets, followed by a Principal Component Analysis.


To measure the possible correlations between Pixel-z\,:\,CAS and QMM, we carried out a regression analysis for the CAS parameters of each physical property against the 8 principal components 
and compared these to a similar
analysis using the CAS parameters for the $r$-band and $u$-band light distributions. The regression analysis shows the extent to which the spatial distribution of physical parameters 
and QMM are related.

For the concentration and asymmetry parameters no correlation was seen.
Thus QMM cannot describe the distribution of physical properties any better than it can describe the $r$-band light distribution for these two parameters. 


The clumpiness of physical parameters is clearly connected better to QMM than is the clumpiness of the r-band or u-band light distributions. This shows that QMM 
(which was developed to quantify morphology) can describe at least some aspects of the distribution of physical properties better than the single filter morphology of a galaxy.
This demonstrates the fact that QMM incorporates physical properties as well as the morphology of galaxies through its inclusion of colour information. 

Variation in redshift did not significantly affect the final
results. The relative correlation between the QMM and physical properties, and QMM and the optical light distribution is consistent across all the redshift bins.
The correlation between the QMM and physical properties, and QMM and the $u$-band light distribution was also consistent across all the redshift bins.

This is a significant step in the process of developing a complete galaxy classification scheme. The QMM approach is a useful technique 
that provides a new way of classifying galaxies. It maybe more closely related to the underlying physical properties of a galaxy than traditional morphology measures. 

This is the first time that there has been an attempt to combine the spatial distribution of physical properties and morphologies of galaxies.
The results are promising, with a definite connection between the spatial distribution of physical properties and the morphology of the galaxies.
This suggests that further investigation is warranted to explore the links between morphology and the underlying physical properties of galaxies.

\section*{Acknowledgements}
We are grateful to the anonymous referee for positive and valuable comments.
DBW acknowledges the support provided by the University of Sydney, School of Physics.
AMH acknowledges support provided by the Australian Research Council
through a QEII Fellowship (DP0557850). BCK acknowledges support from NASA through Hubble Fellowship grant
\#HF-01220.01 awarded by the Space Telescope Science Institute, which is
operated by the Association of Universities for Research in Astronomy,
Inc., for NASA, under contract NAS 5-26555. AJC acknowledges partial support from NSF grants
0851007 and 0709394.

\label{lastpage}

\end{document}